\documentstyle[aps,prl,twocolumn]{revtex}
\begin{document}
\title{Single fluxon in double stacked Josephson junctions: Analytic solution.}
\author{V.M.Krasnov}
\address{Department of Microelectronics and Nanoscience, Chalmers University of \\
Technology, S-41296 G\"oteborg, Sweden \\
and Institute of Solid State Physics, 142432 Chernogolovka, Russia }
\date{\today }
\maketitle

\begin{abstract}
We derive an approximate analytic solution for a single fluxon in a double
stacked Josephson junctions (SJJ's) for arbitrary junction parameters and
coupling strengths. It is shown that the fluxon in a double SJJ's can be
characterized by two components, with different Swihart velocities and
Josephson penetration depths. Using the perturbation theory we find the
second order correction to the solution and analyze its accuracy. Comparison
with direct numerical simulations shows a quantitative agreement between
exact and approximate analytic solutions. It is shown that due to the
presence of two components, the fluxon in SJJ's may have an unusual shape
with an inverted magnetic field in the second junction when the velocity of
the fluxon is approaching the lower Swihart velocity.
\end{abstract}


Magnetic field, $H$, parallel to layers, penetrates stacked Josephson
junctions (SJJ's) in the form of fluxons. Fluxons in SJJ's are different
both from Abricosov vortices in type-II superconductors, since they don't
have a normal core, and from Josephson vortices in a single Josephson
junction (JJ's), since magnetic field is spread over many JJ's. Despite of a
large body of works concerning vortex related properties of layered
superconductors, for review see e.g. Ref. \cite{Blatter}, exact solution for
a single fluxon in SJJ's is still missing. Distribution of magnetic
induction, $B$, at large distances from a fluxon was studied in Refs. \cite
{Clem} and \cite{KrGol}, for tightly packed stacks with identical and
nonidentical layers, respectively. In Ref. \cite{Malomed} a solution for
weekly coupled double SJJ's was obtained using perturbation approach.
However, this solution is not valid for strongly coupled SJJ's, which is the
most interesting case. The electromagnetic coupling is strong when
interlayer space periodicity of the stack, $s$, is (much) less than the
effective London penetration depth, $\lambda _{ab}$, as it is the case for
high-$T_c$ superconductors.

Recently, a simple approximate analytic solution for double SJJ's was
suggested\cite{Modes}. The approximate solution was shown to be in a good
agreement with numerical simulations in the static case for arbitrary
coupling strengths and parameters of the stack. However, the accuracy of the
solution was not strictly proven. Another question which is still opened is
how the shape of the fluxon is changed in the dynamic case. This is a
crucial question since fluxon dynamics determines perpendicular ($c$-axis)
transport properties of SJJ's.

In this paper we rigorously derive the approximate analytic solution for a
single fluxon in a double SJJ's, which is valid for arbitrary junction
parameters and coupling strengths. Using the perturbation theory we find the
second order correction to the solution and analyze its accuracy. Comparison
with direct numerical simulations shows a quantitative agreement between
exact and approximate analytic solutions. It is shown that the fluxon in a
double SJJ's can be characterized by two components, with different Swihart
velocities and Josephson penetration depths. We also studied the
transformation of the fluxon shape with increasing propagation velocities.
It is shown that due to the presence of two components, the fluxon in SJJ's
in the dynamic case may have unusual shape with an inverted magnetic field
in the second junction at high propagation velocities.

We consider a double SJJ's with the overlap geometry, consisting of JJ's 1
and 2 with the following parameters: $J_{ci}$ -the critical current density, 
$t_i$- the thickness of the tunnel barrier between the layers, $d_i$ and $%
\lambda _{Si}$ - the thickness and London penetration depth of
superconducting layers. Hereafter the subscript $i$ on a quantity represents
its number. The strength of electromagnetic coupling of junctions is
determined by the coupling parameter, $S$, which varies from 0 to 0.5 for
the stack with identical JJ's. More details about definitions can be found
in Ref. \cite{Modes}. We consider frictionless fluxon motion with a constant
velocity, $u$. The fluxon will be placed in junction 1.

The physical properties of SJJ's are described by coupled sine-Gordon
equation (CSGE)\cite{Mineev,Bul,SBP} for gauge invariant phase differences, $%
\varphi _i$. The problem with solving CSGE is the coupling of nonlinear $%
sin\left( \varphi _i\right) $ terms. To decouple the variables, we take
linear combination of equation in CSGE and rewrite it as:

\begin{equation}
\widetilde{\lambda }_{1,2}^2F_{1,2\xi \xi }^{\prime \prime }-sin\left(
F_{1,2}\right) =Er_{1,2}\left( \xi \right) ,\   \label{Eq1}
\end{equation}

where $\xi =x-ut$ is the self-coordinate of the fluxon,

\begin{equation}
F_{1,2}=\varphi _1-\kappa _{2,1}\varphi _2,  \label{Eq2}
\end{equation}

\begin{eqnarray}
Er_{1,2} &=&sin(\varphi _1)-\kappa _{2,1}sin(\varphi _2)-sin(\varphi
_1-\kappa _{2,1}\varphi _2)  \nonumber  \label{Eq3} \\
&\simeq &-\kappa _{2,1}\varphi _2\left( 1-cos(\varphi _1)\right) +O\left(
\varphi _2^3\right) .  \label{Eq3}
\end{eqnarray}

Here $\widetilde{\lambda }_{1,2}^2=\lambda _{1,2}^2(1-u^2/c_{1,2}^2)$ and
coefficients $\kappa _{1,2}$, characteristic Josephson penetration depths, $%
\lambda _{1,2}$, and characteristic Swihart velocities, $c_{1,2}$ are given
by Eqs.(17,20,21) from Ref.\cite{Modes}. Such choice for coefficients $%
\kappa _{1,2}$ minimizes $Er_{1,2}$ far from the fluxon center. The phase
differences should satisfy boundary conditions:

\begin{eqnarray}
\varphi _1(-\infty ) &=&0,\varphi _1(0)=\pi ,\varphi _1(+\infty )=2\pi ; 
\nonumber  \label{Eq4} \\
\varphi _2(\pm \infty ,0) &=&0.  \label{Eq(4)}
\end{eqnarray}

Therefore, functions $Er_{1,2}$ have a form of ripple around zero value and
will be considered as perturbation. Solutions of Eq.(1) can be easily found.
Solutions of uniform Eq.(1), i.e. with zero r.h.s., are:

\begin{equation}
F_{1,2}=4arctan\left[ exp\left( \xi /\widetilde{\lambda }_{1,2}\right)
\right] ,  \label{Eq5}
\end{equation}

From Eq.(2) we obtain the first approximation for phase differences:

\begin{eqnarray}
\varphi _1 &=&\frac{\kappa _1F_1-\kappa _2F_2}{\kappa _1-\kappa _2}, 
\eqnum{6 a}  \label{Eq6} \\
\varphi _2 &=&\frac{F_1-F_2}{\kappa _1-\kappa _2},  \eqnum{6 b}
\end{eqnarray}

which coincides with the approximate analytic solution, obtained in Ref. 
\cite{Modes}. The approximate solution is asymptotically correct at large
distances from the fluxon center and has correct values at $x=0$, as follows
from Eqs. (3,4).

Next, we look for a solution of nonuniform Eq.(1) in the form $%
F_{1,2}=F_{1,2}+\delta F_{1,2}$, where $\delta F_{1,2}$ are corrections due
to perturbation terms $Er_{1,2}$:

\begin{equation}
\left( \widetilde{\lambda }_{1,2}^2\frac{d^2}{d\xi ^2}-1+\frac 2{%
cosh^2\left( \xi /\widetilde{\lambda }_{1,2}\right) }\right) \delta
F_{1,2}=Er_{1,2},  \eqnum{7}  \label{Eq7}
\end{equation}

with boundary conditions $\delta F_{1,2}(\pm \infty ,0)=0$. The solution of
Eq. (7) is

\begin{equation}
\delta F_{1,2}=a_{1,2}(\xi )f_{1,2}+b_{1,2}(\xi )g_{1,2},  \eqnum{8}
\label{Eq8}
\end{equation}

where

\begin{eqnarray}
f_{1,2} &=&1/cosh(\xi /\widetilde{\lambda }_{1,2}),  \nonumber \\
g_{1,2} &=&sinh\left( \frac \xi {\widetilde{\lambda }_{1,2}}\right) +\frac 
\xi {\widetilde{\lambda }_{1,2}cosh(\xi /\widetilde{\lambda }_{1,2})}, 
\eqnum{9}  \label{Eq9}
\end{eqnarray}

are partial solutions of the uniform Eq.(7) and

\begin{eqnarray}
a_{1,2} &=&\frac{-1}{2\widetilde{\lambda }_{1,2}}\int_0^\xi Er_{1,2}\left(
x^{\prime }\right) g_{1,2}\left( x^{\prime }\right) dx^{\prime },  \nonumber
\label{Eq10} \\
b_{1,2} &=&\frac 1{2\widetilde{\lambda }_{1,2}}\int_{-\infty }^\xi
Er_{1,2}\left( x^{\prime }\right) f_{1,2}\left( x^{\prime }\right)
dx^{\prime }.  \eqnum{10}  \label{Eq10}
\end{eqnarray}

The perturbation corrections to the approximate fluxon solution, Eq.(6) are

\begin{eqnarray}
\delta \varphi _1 &=&\frac{\kappa _1\delta F_1-\kappa _2\delta F_2}{\kappa
_1-\kappa _2},  \eqnum{11 a}  \label{Eq11} \\
\delta \varphi _2 &=&\frac{\delta F_1-\delta F_2}{\kappa _1-\kappa _2}. 
\eqnum{11 b}
\end{eqnarray}

The corrections $\delta \varphi _{1,2}$ are used to improve the approximate
solution Eq.(6) and to estimate it's accuracy.

Lets observe that from Eq. (3), $\kappa _1Er_1\simeq \kappa _2Er_2$. From
Eqs.(9-11) it can be seen that for $\widetilde{\lambda }_2=\widetilde{%
\lambda }_1$, $\kappa _1\delta F_1=\kappa _2\delta F_2$ and $\delta \varphi
_1\equiv 0$ for arbitrary $\varphi _2\neq 0$. Moreover, taking into account
Eq.(6 b), it can be shown that $d(\delta \varphi _1)/d\widetilde{\lambda }%
_{2(\widetilde{\lambda }_2=\widetilde{\lambda }_1)}=0$.

In Fig. 1, central part of a fluxon is shown for a stack of two identical
JJ's with strong coupling, $S$=0.495, for the static case, $u$=0. Phase
distribution in the full scale is shown in Fig. 3. Parameters of the stack
are: $d_{1-3}=t_{1,2}=0.01\lambda _{J1}$, $\lambda _{S1-3}=0.1\lambda _{J1}$%
, where $\lambda _{J1}$ is the Josephson penetration depth of the single
junction 1. Solid lines represent results of direct numerical integration of
CSGE, Eq. (1), dashed lines show the approximate analytic solution, Eq.(6),
and dotted line shows the corrected analytic solution, $\varphi _{2a}+\delta
\varphi _{2a}$, Eqs.(6,11), in junction 2. Solid and dashed curves, marked
as $\delta \varphi _{1,2}$, represent the overall discrepancy between
numerical and approximate analytic solutions and the perturbation
correction, Eq.(11), respectively.

From Fig. 1 it is seen that correction to the fluxon image, $\varphi _2$, in
the second junction vanishes far from and in the fluxon center, while the
accuracy decreases at distances $\sim \lambda _{J1}$ from the center. Such
behavior is expected from the shape of perturbation functions $Er_{1,2}$ in
Eq. (1). On the other hand, for junction 1, correction $\delta \varphi _1$
is small in the whole space region and analytic solution gives an excellent
fit to the ''exact'' numerical solution, in agreement with discussion above.
The most crucial test for the approximate solution is the accuracy of
derivative at $x=0$

\begin{equation}
\frac{\delta \varphi _1^{\prime }(0)}{\varphi _1^{\prime }(0)}=\frac{\kappa
_1b_1(0)/\widetilde{\lambda }_1-\kappa _2b_2(0)/\widetilde{\lambda }_2}{%
\kappa _1/\widetilde{\lambda }_1-\kappa _2/\widetilde{\lambda }_2}. 
\eqnum{12}  \label{Eq12}
\end{equation}

An estimation for $u=0$ yields

\begin{equation}
\frac{\delta \varphi _1^{\prime }(0)}{\varphi _1^{\prime }(0)}\approx \frac{%
2\alpha \kappa _1\kappa _2\left( \widetilde{\lambda }_1-\widetilde{\lambda }%
_2\right) ^2\left( \widetilde{\lambda }_1^{-1}+\widetilde{\lambda }_2^{-1}+%
\frac{\lambda _0}{\widetilde{\lambda }_1\widetilde{\lambda }_2}\right) }{%
\left( \kappa _1-\kappa _2\right) ^2\left( \widetilde{\lambda }_1+\widetilde{%
\lambda }_2+\frac{2\widetilde{\lambda }_1\widetilde{\lambda }_2}{\lambda _0}%
\right) \left( 1+\frac{\widetilde{\lambda }_1}{\lambda _0}\right) \left( 1+%
\frac{\widetilde{\lambda }_2}{\lambda _0}\right) },  \eqnum{13}  \label{Eq13}
\end{equation}

where $\alpha $ is a factor of the order of unity and $\lambda _0$ given by
Eq.(26) from Ref.\cite{Modes}. From Eq. (13) it is seen that both $\delta
\varphi _1^{\prime }(0)/\varphi _1^{\prime }(0)$ and it's derivative with
respect to $\widetilde{\lambda }_2$ goes to zero at $\widetilde{\lambda }_2=%
\widetilde{\lambda }_1$, as discussed above.

In Fig. 2, the maximum of $\delta \varphi _1\left( x\right) $ (top panel)
and the relative correction to derivative at $x=0$, $\delta \varphi
_1^{\prime }(0)/\varphi _1^{\prime }(0)$, (bottom panel) for $u$=0 are shown
as a function of $J_{c2}/J_{c1}$ for four different coupling parameters $S$%
=0.495 (solid lines), $S$=0.433 (dashed lines), $S$=0.312 (dotted lines) and 
$S$=0.127 (dashed-dotted lines). For $S$=0.495 parameters of the stack are
the same as in Fig.1; for $S$=0.433 $d_i=0.5\lambda _{Si}$; for $S$=0.312 $%
d_i=\lambda _{Si}$; and for $S$=0.127 $d_i=2\lambda _{Si}$. From Fig. 2 it
is seen that the accuracy of solution improves with decreasing $S$. This is
naturally explained by a decrease of $\varphi _2$, see Eq.(3) and decrease
of splitting between $\lambda _{1,2}$, see Eq. (20) from Ref.\cite{Modes}.
However, even for strongly coupled case, the analytic solution, Eq.(6),
gives quantitatively good approximation not only for the value, but also for
the derivative of $\varphi _1$ for arbitrary parameters of the stack. The
gray solid line in the bottom panel of Fig. 2 shows the estimation for $%
\delta \varphi _1^{\prime }(0)/\varphi _1^{\prime }(0)$, calculated from
Eq.(13) for $S$=0.495. It is seen, that estimation gives qualitatively
correct result. Namely, $\delta \varphi _1^{\prime }(0)/\varphi _1^{\prime
}(0)$ vanishes both for $J_{c2}/J_{c1}\rightarrow 0$, as $J_{c2}/J_{c1}$,
and for $J_{c2}/J_{c1}\rightarrow \infty $, as $\sqrt{J_{c1}/J_{c2}}$. Note,
that for $J_{c2}/J_{c1}\rightarrow 0$, $\delta \varphi _1$ vanishes even
though the splitting of $\lambda _{1,2}$ becomes extremely large\cite{Modes}.

Sofar we have considered the static case, $u$=0. On the other hand,
according to Eq. (6), radical changes should take place in the dynamic
state, and the shape of the fluxon in SJJ's may become qualitatively
different from that in the single Josephson junction. Indeed, as the
velocity approaches the lower characteristic velocity, $u\rightarrow c_1$, $%
\widetilde{\lambda }_1\rightarrow 0$, i.e. the $F_1$ component of the fluxon
contracts, while contraction of the $F_2$ component remain marginal. This
implies, that at $u\rightarrow c_1$, the fluxon in SJJ's consists of a
contracted core and uncontracted ''tails'' decaying at distances much larger
than the core size. Such behavior is clearly different from that in a single
Josephson junction. We note that characteristic velocities, $c_{1,2}$ may
depend on $u$, therefore, contraction of each component, $F_{1,2}$, may be
different from Lorentz contraction. Another interesting consequence of the
approximate analytic solution, Eq.(6), is that with increasing $u$, the
magnetic field in the second junction may change the sign with respect to
that in junction 1. Such behavior was predicted in Ref. \cite{Modes} from
the approximate analytic solution, Eq. (6), and it was suggested, that this
will result in attractive fluxon interaction in SJJ's. The magnetic
induction in SJJ's is equal to\cite{Modes}

\begin{eqnarray}
B_1 &=&\frac{H_0\lambda _{J1}}{2\left( 1-S^2\right) }\left[ \varphi
_1^{\prime }+S\sqrt{\frac{\Lambda _2}{\Lambda _1}}\varphi _2^{\prime
}\right] ,  \eqnum{14 a}  \label{Eq14} \\
B_2 &=&\frac{H_0\lambda _{J1}}{2\left( 1-S^2\right) }\left[ S\sqrt{\frac{%
\Lambda _1}{\Lambda _2}}\varphi _1^{\prime }+\frac{\Lambda _1}{\Lambda _2}%
\varphi _2^{\prime }\right] ,  \eqnum{14 b}
\end{eqnarray}
where $H_0=\frac{\Phi _0}{\pi \lambda _{J1}\Lambda _1}$ and $\Lambda _{1,2}$
are defined in Ref. \cite{Modes}.

From Eq.(14) it is seen that estimation of magnetic induction in SJJ's
requires the accuracy of derivatives in both junctions, while Eq.(6) is not
valid with the accuracy of derivative at $x=0$ for $\varphi _2$. Therefore,
more elaborate analysis is needed for the study of magnetic field
distributions in the fluxon.

In Fig. 3, we show a) phase distributions and b) magnetic field
distributions in the fluxon for different fluxon velocities. Parameters of
the stack are the same as in Fig. 1. Solid lines in Fig. 3 a) represent
results of direct numerical simulations of CSGE, Eq.(1), and dotted lines
show the approximate analytic solution, Eq.(6). It is seen that quantitative
agreement between ''exact'' and approximate solutions sustain up to $c_1$.
For the case of identical junctions, considered here, exactly one half of
the fluxon belongs to each of the components, $F_{1,2}$. Indeed, from Fig. 3
a) it is seen that for $u\simeq c_1$ there is a contracted core at $x=0$
with a one $\pi $ step in $\varphi _1$. On both sides of the core, there are
two $\pi /2$ tails, which are slowly decaying at distances $\sim \widetilde{%
\lambda }_2\gg \widetilde{\lambda }_1$. Solid and dashed curves in Fig. 3 b)
represent numerically simulated profiles, $B_{1,2}\left( x\right) $, in
junctions 1 and 2, respectively. From Fig. 3 b) it is clearly seen, that
with increasing fluxon velocity, a dip in $B_2$ develops in the center of
the fluxon. At velocities close to $c_1$, $B_2\left( 0\right) $ changes sign
and finally at $u=c_1$, $B_2\left( 0\right) =-B_1\left( 0\right) $. Such
behavior is in agreement with predictions of Ref. \cite{Modes}. From
numerical simulations we have found, that the fluxon shape in double SJJ's
is well described by Eq.(6) up to at least $u\approx 0.98c_1$ for any
reasonable parameters of the stack, although the accuracy of the approximate
solution may decrease with increasing $u$. The decrease of accuracy is
caused by the increase of $\varphi _2$ as is seen from Fig. 3. In this case,
perturbation correction, Eq. (11), should be taken into account.

In conclusion, a simple approximate analytic solution for a single fluxon in
a double stacked Josephson junctions for arbitrary junction parameters and
coupling strengths is derived. It is shown that the fluxon in a double SJJs
can be characterized by two components, with different Swihart velocities
and Josephson penetration depths. Using the perturbation theory we find the
second order correction to the solution and analyze its accuracy. Comparison
with direct numerical simulations shows a quantitative agreement between
exact and approximate analytic solutions for all studied parameters of the
stack and fluxon velocities up to at least 0.98$c_1$. It is shown that due
to the presence of two components, the fluxon in SJJ's may have an unusual
shape with an inverted magnetic field in the second junction at large
propagation velocities. This may lead to attractive fluxon interaction in
the dynamic state of SJJ's

Discussions with D.Winkler are gratefully acknowledged. The work was
supported by the Russian Foundation for Basic Research under Grant No.
96-02-19319.

\begin{figure}[tbp]
\caption{ Central part of a fluxon in a stack of two identical JJ's with
strong coupling, $S$=0.495, for the static case, $u$=0. Solid and dashed
lines represent "exact" numerical and approximate analytic solutions,
respectively. Dotted line shows the corrected analytic solution, $\varphi
_{2a}+\delta \varphi _{2a}$.}
\label{Fig. 1}
\end{figure}

\begin{figure}[tbp]
\caption{ Perturbation correction to the approximate analytic solution is
shown as a function of $J_{c2}/J_{c1}$ for the maximum of $\delta \varphi
_1\left( x\right) $ (top panel) and the relative correction to derivative at 
$x=0$, $\delta \varphi _1^{\prime }(0)/\varphi _1^{\prime }(0)$, (bottom
panel) for $u$=0 and for four different coupling parameters. The gray solid
line in the bottom panel shows the estimation from Eq.(13) for $S$=0.495. It
is seen that the analytic solution, Eq.(6), gives quantitatively good
approximation not only for the value, but also for the derivative of $%
\varphi _1$ for arbitrary parameters of the stack. }
\label{Fig. 2}
\end{figure}

\begin{figure}[tbp]
\caption{ Fluxon shapes in double SJJ's for different fluxon velocities, $%
u/c_1$=0, 0.61, 0.92, 0.98, 0.998, 0.9999 (from left to right curve). In
Fig. 1 a) phase differences, $\varphi _{1,2}$, are shown. Solid and dotted
lines represent ''exact'' numerical and approximate analytic solutions
respectively. Fig.1 b) shows magnetic inductions $B_{1,2}$ obtained
numerically. The existence of contracted and uncontracted components and the
sign inversion of $B_2(0)$ at $u\simeq c_1$ is clearly seen.}
\label{Fig. 3}
\end{figure}

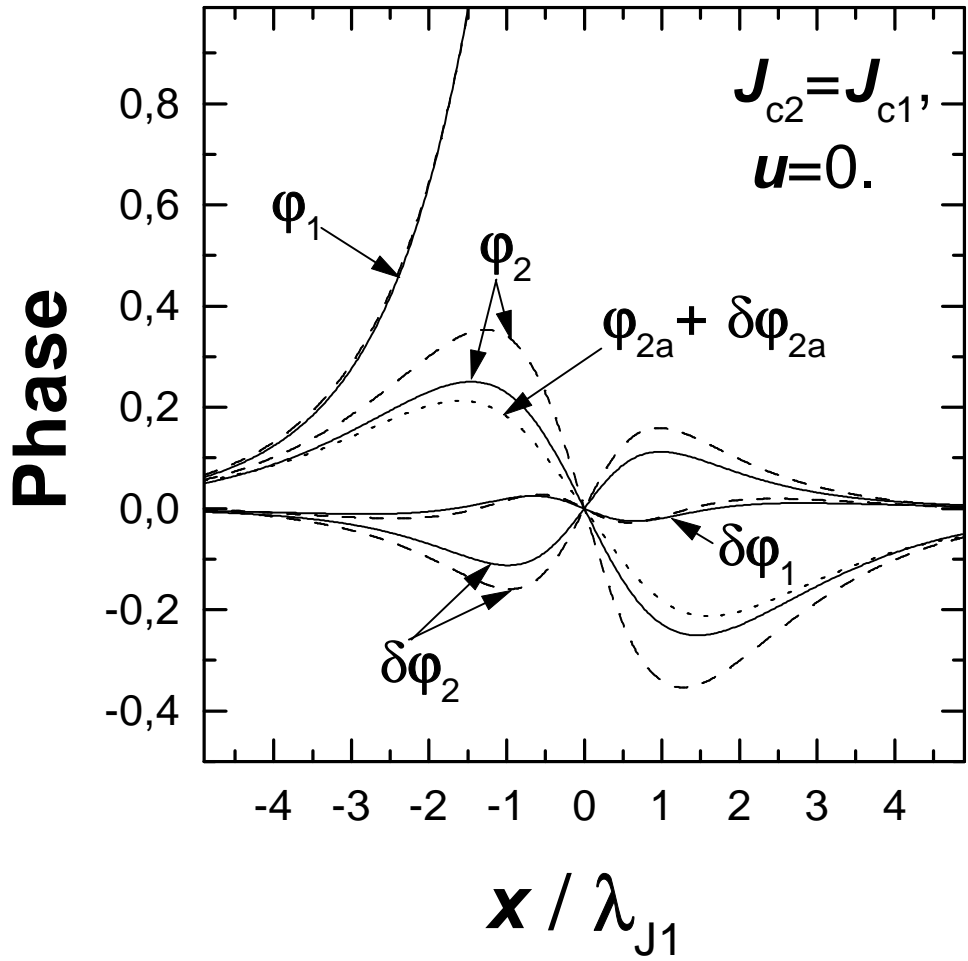
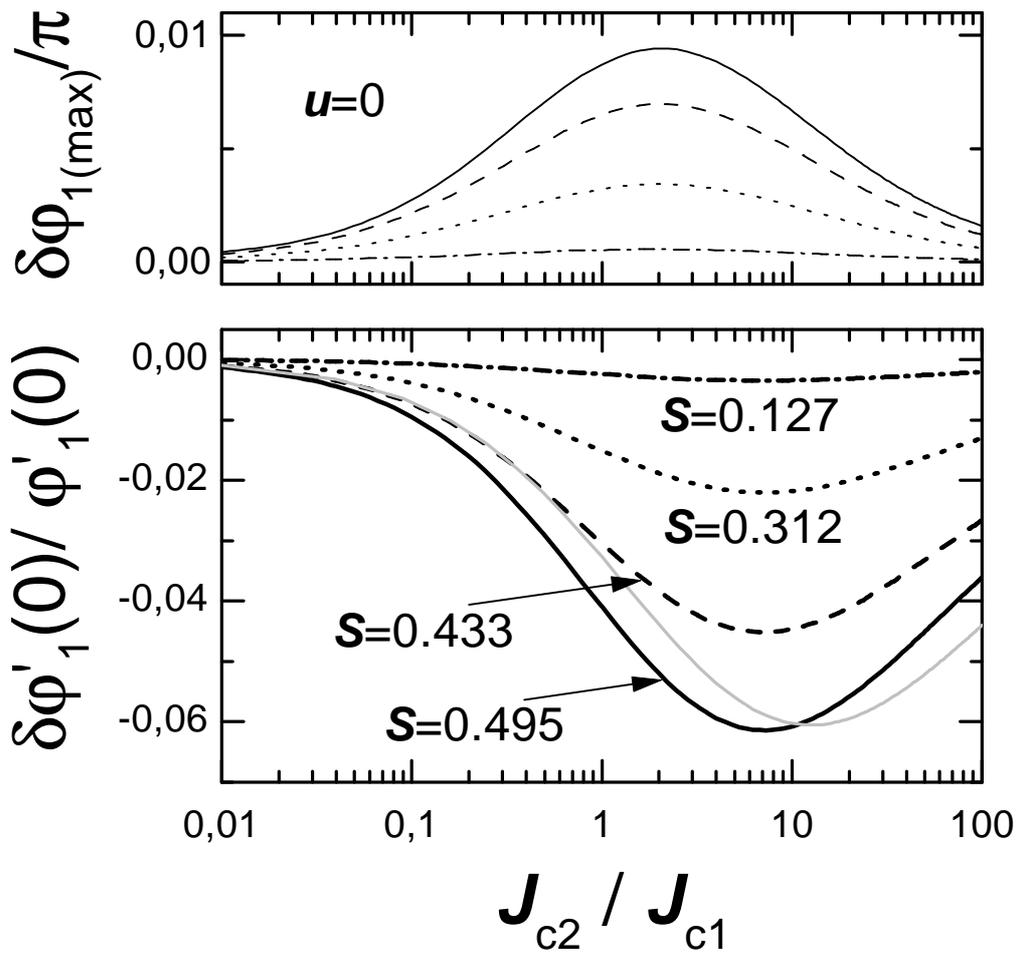
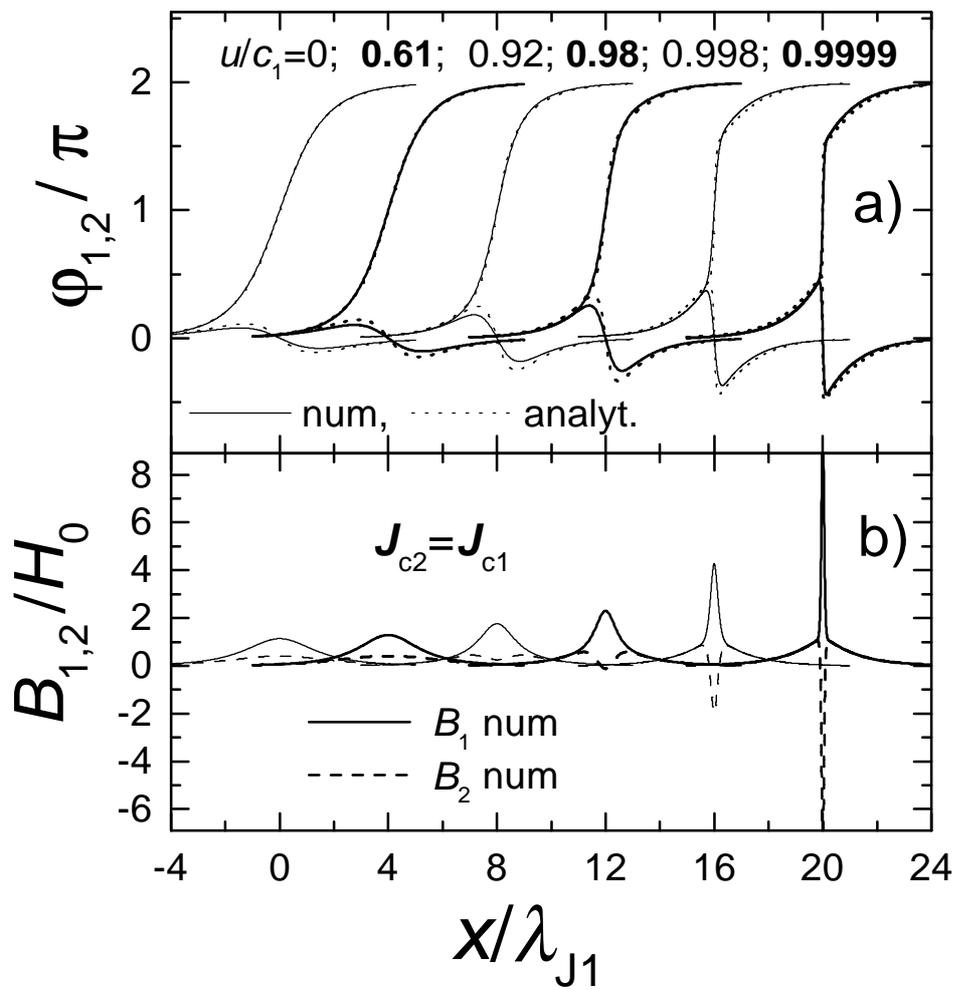
\end{document}